# Enhancement of superconductivity in NbN nanowires by negative electron-beam lithography with positive resist

I. Charaev,[1, a)] T. Silbernagel,[1] B. Bachowsky,[1] A. Kuzmin,[1] S. Doerner,[1] K. Ilin,[1] A. Semenov,[2] D. Roditchev,[3] D. Yu. Vodolazov,[4] and M. Siegel[1]

[1]*Institute of Micro- und Nanoelectronic Systems, Karlsruhe Institute of Technology (KIT), Hertzstrasse 16, 76187 Karlsruhe, Germany*
[2]*Institute of Optical Systems, German Aerospace Center (DLR), Rutherfordstrasse 2, 12489 Berlin, Germany*
[3]*Institut des Nanosciences de Paris, Université Pierre et Marie Curie-Paris 6 and CNRS-UMR 7588, 4 place Jussieu, 75252 Paris, France*
[4]*Institute of Physics of Microstructures, Russian Academy of Sciences, 603950 Nizhny Novgorod, GSP-105, Russia*

We performed comparative experimental investigation of superconducting NbN nanowires which were prepared by means of positive-and negative electron-beam lithography with the same positive tone Poly-methyl-methacrylate (PMMA) resist. We show that nanowires with a thickness 4.9 nm and widths less than 100 nm demonstrate at 4.2 K higher critical temperature and higher density of critical and retrapping currents when they are prepared by negative lithography. Also the ratio of the experimental critical-current to the depairing critical current is larger for nanowires prepared by negative lithography. We associate the observed enhancement of superconducting properties with the difference in the degree of damage that nanowire edges sustain in the lithographic process. A whole range of advantages which is offered by the negative lithography with positive PMMA resist ensures high potential of this technology for improving performance metrics of superconducting nanowire singe-photon detectors.

## I.  Introduction

During last 15 years the technology of superconducting nanowire single-photon detectors (SNSPDs) is under intense development, continuously improving SNSPDs performance. Many efforts in the fields of optics, solid state physics and thin-film technology have been attempted in order to find the way to increase detection efficiency (DE) at longer wavelengths and to decrease timing jitter and dark count rate (DCR) of such detectors. The development of SNSPDs follows several main directions. The search for optimal materials and the improvement of the quality of superconducting nanowires are among the most straightforward ones.

First SNSPDs were made from thin NbN [1] and NbTiN [2] films. These materials can be reliably elaborated and patterned; they have hard superconducting gap and critical temperatures well above the temperature of liquid helium that facilitates their use. Though, they exhibit a roll-off in the wavelength dependence of the detection efficiency which begins at wavelengths $\lambda \leq 1$ μm. As compared to nitrides, uniform thin films of amorphous superconductors like WSi [3], NbSi [4], MoGe [5] and MoSi [6] were shown to be

---

[a)] Electronic mail: ilya.charaev@kit.edu

more promising for an effective detection of near infrared photons with larger wavelengths. The drawback is a noticeably lower critical temperature and correspondently smaller energy gaps, which unavoidably results in lower critical current densities. Moreover, it has been found [5, 6] that SNSPDs from these low-temperature materials exhibit larger timing jitter than those from nitrides [7]. That is why high-quality ultrathin nitride films remain a reference in a highly competitive field of SNSPDs.

It has been confirmed experimentally that reduction of the cross-section of nanowires is a good approach for enhancement of the DE of SNSPD for photons with small energies [8, 9, 10]. However, this approach has several limitations. Ultra-thin films undergo superconductor-to-insulator transition [11], they are characterized by a reduced absorbance [9] and by a stronger spatial non-uniformity of the superconducting energy gap [12]. The decrease in the cross-section of nanowires also results in a smaller experimental critical current $I_C$ which causes a relatively low amplitude-to-noise ratio (ANR) and significant timing jitter in the voltage transients after the photon absorption events [13]. Furthermore, thin and narrow nanowires require special efforts to optimize optical absorption in meanders of typical SNSPDs [14]. Theoretical models [15, 16] of photon detection in SNSPDs predict an increase of a cut-off wavelength $\lambda_0$ of DE($\lambda$)-dependence when the current applied to the nanowire approaches the depairing current $I_C^{dep}$. In practice, SNSPDs operate at a bias current which is slightly less than the experimentally achievable critical current $I_C$. The latter is usually smaller or even much smaller the departing current. Hence, the useful spectral bandwidth of SNSPD could in principle be extended to larger wavelengths via pushing $I_C$ towards the depairing current limit $I_C = I_C^{dep}$. This idea is very attractive. Once the current ratio $I_C/I_C^{dep}$ is enhanced, it would enable operation of detectors (made of materials with high $T_C$) at relatively large temperatures and high bias currents. There have been already several reports on possible approaches towards enhancement of $I_C/I_C^{dep}$ ratio. One of the reasons why the ratio remains significantly below unity is the crowding of supercurrent in the vicinity of sharp bends inherent to nanowires in the form of a meander [17]. The current crowding can be effectivity suppressed by optimization of the detector layout in that a radius of the bends plays the major role. Increase of the bending radius decreases the strength of current crowding, thus lowering both the dark count rate and timing jitter [18, 19, 20]. It has been also shown that the adjustment of stoichiometry of NbN films towards higher Nb-content results in an increased $I_C/I_C^{dep}$ ratio and thereby in a broadening of the spectral bandwidth of SNSPDs [21]. In general, detection efficiency is higher for SNSPDs with local values of the $I_C/I_C^{dep}$ ratio uniformly distributed over the nanowires and bends. This ratio is affected by different types of defects such as cross-section variations due to the non-uniformity of the thickness or width of the film, nanowire edge defects or internal structural defects weakening the superconducting order.

All those reduce the current ratio and make detector characteristics worse. Therefore, development and optimization of approaches, which target growth of highly uniform thin films and pattering them in defect-free superconducting nanowires, are in permanent focus of numerous research groups. In this paper we focus on the effect of patterning on superconducting properties of NbN nanowire.

Usually the nanowires below 100 nm in width are obtained out of thin superconducting films by electron-beam lithography. The films are spin-coated with an electron-beam resist (PMMA, HSQ, ZEP among others), which is then locally affected by the electron beam. After the lithography, unprotected parts of the film are subsequently removed by one of available etching techniques. Each step of the process may potentially influence the SNSPD performance. For instance, the choice of the resist type and the spin-coating procedure decide the smallest pixel-size which is possible to write and, consequently, the nanowire edge roughness. Ideally, to achieve the ultimate $I_C/I_C^{dep}$ ratio for a specific layout of SNSPD, the size of defects should be reduced below the coherence length, in order to do not affect the critical current. Since the lithography resolution is usually improved with reducing the thickness of the resist layer, the straightforward tendency would be to make it as thin as possible. However, the stability of the resist during etching should be sufficient to prevent the etching of the film under the protecting resist layer and at its edges. Thinner layers are more fragile against the etching attack, leading to rougher nanowire edges.

Here we demonstrate that via strong overexposure the standard (positive tone) PMMA electron-beam resist can be made readily suitable for the negative lithographic process and that this procedure enables significant improvement of superconducting characteristics of NbN nanowires. The nanowires with a width less than 100 nm demonstrate enhanced superconducting critical temperature, densities of the critical and retrapping currents as well as the $I_C/I_C^{dep}$ ratio. We relate this enhancement to reduced non-perfection of nanowire edges and invoke theoretical considerations [15, 16] to estimate expected improvement in SNSPD performance which will be introduced by the negative-PMMA lithography.

## II. Technology

### A. Thin-film deposition

Thin NbN films have been deposited simultaneously on two identical 10x10 mm$^2$ single-side polished substrates from R-plane-cut sapphire via reactive magnetron sputtering of pure Nb target in an atmosphere of mixed argon and nitrogen gases. Partial pressures of argon and nitrogen were $P_{Ar} = 1.9 \times 10^{-3}$ mbar and $P_{N2} = 3.9 \times 10^{-4}$ mbar, correspondingly. The substrates were placed without been thermally anchored on the surface of a copper holder, which was in turn placed onto a heater plate. During the deposition of NbN layer the plate was kept at a temperature of 850°C. The deposition rate of NbN was 0.14 nm/s at the discharge current of 275 mA. These conditions ensure the stoichiometry of NbN films which results in the highest critical temperature for a given thickness. The film thickness $d = 4.9\pm0.2$ nm was measured by a stylus profiler.

### B. Nanowire patterning

The films were patterned into nanowires via the electron-beam lithography over the PMMA resist and subsequent Ar ion milling.

The PMMA resist is a well-known positive-tone resist which is attractive for users due to easy handling, high temporal stability, reproducibility of lithographed structures, and high-resolution. The PMMA resist is available with different sensitivities. The required thickness of the resist layer can be easily achieved by varying the speed of spinning and/or the amount of solid content in the resist. The PMMA itself and the required developer and stopper are water-free materials; this is preferable for pattering of films from water-sensitive materials. Certain disadvantage of the PMMA resist is a relatively high temperature (between 150 and 190°C), which is required to bake the resist after spinning. Such high temperature stimulates diffusion of oxygen that increases its penetration depth into the film where, for films from Nb compounds, oxygen deteriorates or suppresses superconductivity. Furthermore, this resist is moderately stable against plasma assisted etching processes. This limits its applicability especially in the case of thin layers, which are required for writing ultimately small features. Although the PMMA electron-beam resist was originally introduced as a positive-tone resist, it can be used for negative lithographic processes also.

When a primary beam of electrons with energies ≈10 keV (far below the threshold for the displacement of the carbon atom [22]) enters the PMMA resist and a substrate, it produces low-energy secondary electrons (SE), which are mainly responsible for the scission process of the PMMA polymer chain [23]. A reduced molecular weight of the PMMA resist, which is exposed with a dose in the range of 100 μC/cm$^2$, makes it soluble in solvents with the high enough activation energy [24]. In this case PMMA acts as a positive-tone resist.

At high exposure doses ($\geq$ 1 mC/cm$^2$) PMMA chains decompose into very short low-molecular-weight fragments, which start to form a dense carbonized film. The structures made of this film are insoluble in a standard PMMA-developer and even in acetone due to a cross-linking and formation of the covalent bonds between fragments. In this case PMMA acts as a high-resolution negative-tone electron-beam resist [25].

Two identical NbN films were patterned simultaneously one by the positive and another by the negative process in order to eliminate different aging degrees of the films. The substrates were spin-coated with PMMA 950k resist with a layer thickness of 95 nm. In order to minimize degradation of the films, the resist was baked on a hot plate at the lowest recommended temperature of 150°C for 5 minutes.

Layout was the same for all samples in both the positive- and negative-PMMA series and represented a straight nanowire with a typical for SNSPDs width $W \leq 100$ nm, which was embedded between small contact pads. In order to avoid the current crowding at steps from the pads with a width of a few tens of micrometers to the nanowire, they were rounded off with a radius $r \approx 4$ μm. In each series, the width of nanowires was varied between 50 and 100 nm. Additionally, several strips with similar layout but with a width of a few micrometers were made to serve as reference structures. We assumed that the influence of edges is negligible for such wide strips and their properties are mostly determined by the patterning and aging. The actual widths of nanowires and strips were measured using the scanning electron microscopy (SEM).

In the case of "standard" positive-PMMA process, two separate islands were exposed by 10 kV electron beam with the dose about 100 μC/cm$^2$. The islands were separated by a slit which in its middle had the width equal to the design width of the nanowire and grew at edges to encompass rounded steps to contact pads. After development in the standard developer for 30 sec (30% MIBK in 2-propanol at 23 °C) and rinsing in the 2-propanol stopper, the exposed areas were removed. Unexposed resist between islands remained on the surface of film (Fig. 1a) and protected the film during the subsequent ion-milling process. Large contact pads sized to a few millimeters for ultrasonic bonding were prepared by photolithography with a mask which additionally protected the already patterned nanowire and small pads during the second etching step and separated samples from each other making them ready to measure.

In the negative-PMMA process, exposure dose of the resist was increased by two orders of magnitude to reach 10 mC/cm$^2$ while the energy of electrons was kept unchanged. In contrast to the positive-PMMA process described above, here the electron beam exposed only nanowire and small pads (Fig. 1b) and after development these areas remained on the film surface. The pattern was developed in acetone for 1.5 min and then rinsed in 2-propanol. Right after that, large contact pads were prepared by standard photolithography with a mask which left area with the nanowire and rounded steps under negative-PMMA open but overlapped with the small pads. After development of the photoresist, the complete image containing the central part with negative-PMMA and the large contact pads was transferred into NbN film by ion-milling process.

We note that the thickness of the PMMA resist at the areas, which were exposed with the large dose of 10 mC/cm$^2$, shrunk from 95 nm to about 50 nm (this value was measured right after exposure) and remained unchanged after the development in acetone. No measurable changes were observed in the thickness of PMMA resist exposed by the low dose of 100 μC/cm$^2$. Etching rate of the negative-PMMA resist was found to be about 2.7 nm/min. This rate is comparable to the etching rate of NbN by Ar ions with the energy 200 eV and current density ≈ 1 mA/cm$^2$ at 10° incident angle. At the same etching conditions, the etching rate of the positive-PMMA resist was almost 2.5 times higher (≈ 6.7 nm/min). After the etching a residual resist (both positive and negative) was removed from the surface of NbN film using a combination of the warm acetone, ultrasonic shaking and a gentle mechanical brushing. The possibility to remove a hardened PMMA mask after etching is essential for multi-layer structures such as a single-spiral SNSPD [20].

However, it has to be noted that the increase of the exposure dose by two orders of magnitude and the increase of the area, which has to be exposed in the case of negative-PMMA lithography, result in increased writing time with the electron beam. In turn, the increase in the writing time requires additional efforts to get long-term stability of electron-beam parameters and long-term suppression of external acoustic, mechanic, and electro-magnetic interferences disturbing the lithography apparatus.

### III. Results of experimental characterization

#### A. NbN film

The temperature dependence of the square resistance $R_{sq}$ of the NbN film was measured immediately after deposition at temperatures from 4.2 up to 300 K by the standard four-probe technique. The critical temperature of the film was 13.55 K. The specific resistivity $\rho$ was evaluated using the measured thickness $d$ and the square resistance of the film as $\rho = R_{sq} \times d$. The residual resistivity $\rho_0$ at $T = 25$ K was about 120 μΩ×cm. The residual-resistivity ratio ($RRR$) of the film, i.e. the ratio of the resistivity at room temperature to the residual resistivity, was slightly larger than one ($RRR \approx 1.02$). The temperature dependence of the second critical magnetic field $B_{C2}(T)$ was measured at temperatures in the vicinity of the transition in an external magnetic field up to 3 T applied perpendicular to the film surface. The value of the second critical magnetic field at zero temperature was calculated in the dirty limit [26] and amounted at $B_{C2}(0) = 18.8$ T. With this value the coherence length at zero temperature was calculated as

$$\xi(0) = \sqrt{\frac{\Phi_0}{2\pi B_{C2}(0)}} \quad . \tag{1}$$

The found value $\xi(0) \approx 4.2$ nm is close to the thickness of our film. The magnetic field penetration depth was calculated as

$$\lambda(0) = \sqrt{\frac{\hbar \rho_o}{\pi \mu_0 \Delta(0)}} \tag{2}$$

where $\Delta(0) = 2.05 k_B T_C$ is the superconducting energy gap of NbN [27, 12]. We found $\lambda(0) = 287$ nm which is close to the value [11] obtained by the inductive technique for similar NbN films.

**B. NbN nanowires**

After patterning, the resistance of all nanowires was measured from the room temperature down to 4.2 K. Critical temperatures of nanowires were lower than those of the micrometer-wide reference strips made from the same film. Independently of the type of lithographic process, the $T_C$ of the micrometer wide strips was almost equal to the critical temperature of the non-patterned film. It is seen in Fig. 2 that the critical temperature $T_C$ of the nanowire decreases with decreasing width. For the same interval of widths between 50 and 100 nm, $T_C$ of the positive-PMMA nanowires varies in the range between 11.4 and 12.5 K while $T_C$ of the negative-PMMA nanowires varies from 12 to 12.8 K and is in average 0.5 K higher.

Current-voltage characteristics of all nanowires were measured at $T = 4.2$ K in the current bias mode. The critical current $I_C$ is associated with the step-like switching of the nanowire from the superconducting to the normal state when the current increases from zero. The nominal density of the critical current at $T = 4.2$ K was calculated as $j_C(4.2\text{ K}) = I_C(4.2\text{ K})/Wd$. The graph in Fig. 3 represents the dependence of $j_C(4.2\text{ K})$ on the width for the positive-PMMA (black squares) and the negative-PMMA (red circles) nanowires. Similar to the critical temperature, the critical current density in both series decreases with the decreasing width of nanowires. It is seen that the density of the critical current of the negative-PMMA nanowires, which is in the range $15 \div 17$ MA/cm$^2$, is in average 40% larger than the $j_C(4.2\text{ K})$ values of the positive-PMMA nanowires ($10 \div 14$ MA/cm$^2$).

The nominal density of the retrapping current was calculated as $j_r(4.2\text{K}) = I_r(4.2\text{ K})/Wd$. Here $I_r(4.2\text{ K})$ is the value of the bias current at which the nanowire returns from the resistive to the superconducting state when the bias current decreases. The obtained results are shown in Fig. 4. The $j_r(4.2\text{ K})$ of the positive-PMMA nanowires (black squares) increases from 2.8 up to 3.1 MA/cm$^2$ with the decreasing width. The values of $j_r$(4.2 K) of the negative-PMMA nanowires (red circles) are higher ($\approx 3.3 \div 3.8$ MA/cm$^2$) but the $j_r$(4.2 K)-dependence on the width does not have any distinct behavior. In the case of 50 nm wide nanowires, the difference in the density of the retrapping current between two series is about 20%.

## IV.  Discussion

The negative-PMMA lithography offers the following advantages over the positive-PMMA lithography. The negative-PMMA resist is more stable which is seen in more than two times smaller etching rate. Furthermore, since the critical temperatures of micrometer wide reference strips made by the positive- and the negative-processes were the same, the 50 nm thick layer of the negative-PMMA resist protects the film during etching process as good as almost the twice thicker layer of the positive-PMMA resist. In the electron-beam lithography, smaller thickness of the resist layer, in general, allows for writing smaller pixels in a more reproducible manner. Therefore, further reduction of the width of nanowires is possible with the negative-PMMA process while keeping the protection properties of the resist at high enough level. This is already seen in Fig. 2 where the critical temperature of the 47 nm wide nanowire made by the negative-PMMA process is as high as the $T_C$ value of the 80 nm wide nanowire from the positive-PMMA series. While the enhancement of the critical temperature of negative-PMMA nanowires is about 0.5 K, which is only 5% of the $T_C$ of the positive-PMMA nanowires, the increase in the critical current density measured at 4.2 K is much more significant (Fig. 3). The latter makes the negative-PMMA technology even more attractive for optimization of SNSPDs. Larger critical current, which can be realized for the given width, means not only larger amplitude-to-noise ratio (ANR) in the voltage-pulse response of the detector but, according to [13], also leads to smaller values of the timing jitter. The higher retrapping current density (Fig. 4) can be interpreted as a result of enhanced cooling efficiency for the negative-PMMA nanowires. For SNSPDs, this should lead to lower dark count rates [28] and to smaller latching probability.

Higher $T_C$ corresponds to a larger energy gap. With other material parameter and operation conditions being equal, this should reduce a value of the cut-off wavelength. From the other side, the less $T_C$ of nanowires is reduced the thinner films with lower critical temperature (the critical temperature decreases with decreasing thickness due to the proximity effect [29]) can be used to fabricate SNSPD for operation at a given temperature, e.g. at $T = 4.2$ K. Furthermore, higher $T_C$ is one of the reasons for higher $j_C$ of the negative-PMMA nanowires. However, increase in $T_C$ alone cannot explain almost 40% increase in $j_C$(4.2 K). In Fig. 5 the ratio of the densities of the measured critical current and the de-pairing critical current at 4.2 K $j_C/j_C^{dep}$(4.2K) is shown for all nanowires. The densities of the depairing critical current were computed with Eq. 3 for the actual $T_C$ of each nanowire (Fig. 2). The temperature dependence of $j_C^{dep}$ was adopted from the work of Kupriyanov and Lukichev [30]. We used the $KL(T)$-correction (Fig. 1 in Ref. 30) for the extreme dirty limit to the Ginzburg-Landau (GL) departing current density and obtained $j_C^{dep}$ for our films as:

$$j_C^{dep}(T) = \frac{16\sqrt{\pi}\exp(2\gamma)}{21\varsigma(3)\sqrt{6}} \beta_0^2 \frac{(k_B T_C)^{\frac{3}{2}}}{e\rho\sqrt{D\hbar}} \left(1 - \frac{T}{T_C}\right)^{\frac{3}{2}} KL(T) . \quad (3)$$

In Eq. (3) $e$, $\hbar$, $k_B$ are the fundamental constants, $\gamma = 0.577$, $\zeta(3) = 1.202$, $\beta_0 = \Delta(0)/(k_B T_C) = 2.05$ [12, 27] and $D = 0.56$ cm$^2$/s is the electron diffusivity in NbN film which was calculated as

$$D = -\frac{4k_B}{\pi e} \left(\frac{dB_{c2}}{dT}\right)^{-1}_{T \to T_C} . \quad (4)$$

Here $dB_{C2}/dT$ is the temperature derivative of the second critical magnetic field in the vicinity of the critical temperature. It is seen that the $j_C/j_C^{dep}$(4.2K) ratio for the negative-PMMA nanowires is approximately 0.67 ($\pm$ 3) that is noticeably larger than the value 0.48 ($\pm$ 5) obtained for the positive-PMMA nanowires. According to the theoretical models of SNSPD response [15, 16] an increase of this ratio should result in the shift of the cut-off wavelength towards longer wavelengths.

Although the superconductivity in the negative-PMMA nanowires is enhanced the ratio $j_C/j_C^{dep}$(4.2 K) is still less than one. One of the possible reasons for that can be the particular stoichiometry of our NbN films. The films were deposited at the discharge parameters providing maximum of $T_C$. Larger $j_C/j_C^{dep}$(4.2 K) ratio is achieved in films with larger relative content of Nb [21] which, however, have slightly smaller critical temperature. Another reason is revealed by the temperature dependence of the critical current density of the 80 nm wide nanowire which is shown in Fig. 6. Such dependence is typical for nanowires with $W \leq 100$ nm. As function of reduced temperature in the form $t = (1-T/T_C)^{3/2}$, $j_C$ increases linearly (the dashed line) at temperatures in the vicinity of $T_C$. At $t \approx 0.08$ the experimental points deviate from the linear dependence. The deviation becomes larger at temperatures close to 4.2 K. Although the linear part is present in the $j_C(t)$ dependences for nanowires from both series, its slope and the temperature at which the deviations begins vary from sample to sample affecting the current ratio $j_C/j_C^{dep}$(4.2 K). Physical phenomena, which cause this deviation and variations of the slope, are out of the scope of the present work. They will be considered in details elsewhere [31].

Here we would like to deal only with the high temperature part of the $j_C(t)$ dependence near $T_C$ ($t \to 0$) where the critical current density increases linearly with the reduced temperature. Since the width of our nanowires is much smaller than the Pearl length

$$\Lambda = \frac{2\lambda^2}{d}, \tag{5}$$

the supercurrent is assumed to be distributed evenly over the cross-section of the nanowire. With this approximation, the critical current of nanowires should be determined by the de-pairing mechanism solely at least at temperatures close to $T_C$ where the film thickness is additionally much less than the coherence length (1). In the framework of this approximation, the slope of the linear fit of $j_C(T)$ dependence at $t \ll 1$, $j_C^{extr}$ (the dashed line in Fig. 6), should equal the temperature independent coefficient $j_C^{dep}(0)$ in the KL-corrected Ginzburg-Landau expression (3) for the temperature dependence of the density of the depairing current. Since at low temperatures the true depairing current (3) drops below the Ginzburg-Landau depairing current, $j_C^{extr}$ can be seen as the parameter characterizing the ultimate current-carrying ability of a particular nanowire. From the measured $j_C(T)$-dependencies we found $j_C^{extr}$-values for all nanowires. The dependences of the ratio $j_C^{extr}/j_C^{dep}(0)$ on the width of the positive- and negative-PMMA nanowires are shown in Fig. 7. In the case of the negative-PMMA nanowires, this ratio (in average) amounts at 0.85 (still less than one) while for the positive-PMMA nanowires it is approximately 0.58.

The results presented above evidence the potential of the negative PMMA resist for further development and optimization of SNSPDs which are targeted at high detection efficiency for low energy photons. This potential, however, can be fully realized only if the reason for the deviation of $j_C$ from the density of the depairing critical current will be eliminated.

From the technological point of view the advances of the negative-PMMA nanowires over the positive-PMMA nanowires can be understood with the assumption that edges of nanowires are less damaged in the negative process. Consequently, the negative-PMMA nanowires have larger effective superconducting cross-section. This leads to a weaker proximity effect and to increased $T_C$ (Fig. 2). Larger superconducting cross-section of the nanowire is capable to carry larger critical current. Therefore, for a given nominal width, the calculated density of the critical current is also larger for the negative-PMMA nanowires (Fig. 3). The same arguments explain the increase of the retrapping current density (Fig. 4).

To prove the assumption that edges of the nanowire are less damaged in the negative process, we performed statistical analysis of the edge roughness of our nanowires. In Fig. 1 there are two SEM images of nominally 100 nm wide nanowires after ion milling. Nanowires were prepared by the positive (Fig. 1c) and negative (Fig. 1d) electron-beam lithography of PMMA resist. The rest of resist was stripped off and samples were thoroughly cleaned. For the analysis of the nanowire geometry, we used a high-resolution field-emission SEM equipped with the highly-efficient in-lens detector of secondary electrons (SE). This detector is sensitive mainly to secondary electrons, which are generated in the vicinity of the primary beam at the very surface of a sample ("SE type-I"). Therefore, the in-lens detector ensures spatial resolution of about 1-2 nm. The raster scan was aligned along the length of nanowire to minimize low-frequency noise contribution to the image. The set of images with different magnification were acquired with the nominal resolutions better than 0.5 nm/pixel.

We used data from the SEM images to extract the mean value and the standard deviation of the width of nanowires fabricated by the negative- and positive-PMMA lithography. In order to do that, we used the recorded raster scan which is 2D-array with 255 gradations of the SE-signal intensity per pixel. The lines, which correspond to the edges of the nanowire, were defined using the Canny edge-detection algorithm. The local width of the nanowire is defined as a distance between detected edges in the direction perpendicular to the line of the raster scan. The standard deviation was obtained from the Gaussian fit of the histogram of the local wire-width. The standard deviation for the positive-PMMA nanowires is $\approx$ 5 nm, while for the negative-PMMA nanowires it is smaller $\approx$ 2-3 nm. There are several factors which may result in increased edge roughness of the positive-PMMA nanowires. The presence of the solvent residues in the positive PMMA can lead to the non-uniform development process. The etching rate of the PMMA by accelerated Ar ions could be also different for the areas with the trapped solvent residues. During ion milling, irradiation by the low energy Ar ions could lead to carbonization of the resist layer. Consequently, structures from the positive PMMA will contract non-uniformly. This can also lead to the edge damage of the nanowire.

### V. Conclusions

We investigated experimentally and then compared superconducting properties of thin-film NbN nanowires which were fabricated using the positive and negative electron-beam lithography over standard positive PMMA resist. Nanowires with smaller width and smaller edge roughness were obtained in a reproducible manner due to higher resolution of the negative-PMMA lithography process. The higher critical temperature (about 5% enhancement) of the negative-PMMA nanowires allows further decreasing the cross-section of the nanowire while keeping $T_C$ high enough for operation of SNSPD at 4.2 K. The improved cooling efficiency of the negative-PMMA nanowires manifests itself as the 20% higher values of the retrapping current density. For SNSPDs, increase in the cooling efficiency should result in lower dark count rate and lower latching probability. Furthermore, approximately 30% higher critical current density in the negative-PMMA nanowires should improve signal to noise ratio of detectors, while the 40% higher $j_C/j_C^{dep}$ ratio at 4.2 K should shift the cut-off wavelength of NbN SNSPD's response towards the infrared range. Critical current density $j_C = 0.85 j_C^{dep}$ in the negative-PMMA nanowires should be reachable also at the operation temperature once edge quality of the nanowires further improves.

**ACKNOWLEDGMENT**

I.C. acknowledges support from Karlsruhe School of Optics and Photonics of Karlsruhe Institute of Technology. The work was supported in part by the DFG/ANR project SUPERSTRIPES SI 704/11-1. D. Yu. V. acknowledges support from Russian Foundation for Basic Research (project No 15-42-02365).

**Figures**

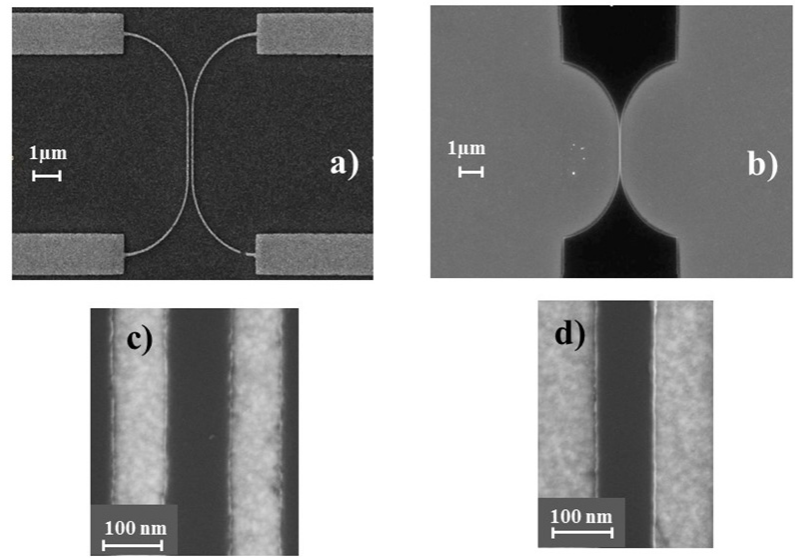

FIG. 1 SEM images of nanowires which were prepared by the positive-PMMA (a, c) and the negative-PMMA (b, d) lithography: a) and b) – after development of the resist (dark areas – PMMA resist); c) and d) – central parts of nanowires prepared by two different lithography after ion milling and stripping off the resist (dark areas – NbN film).

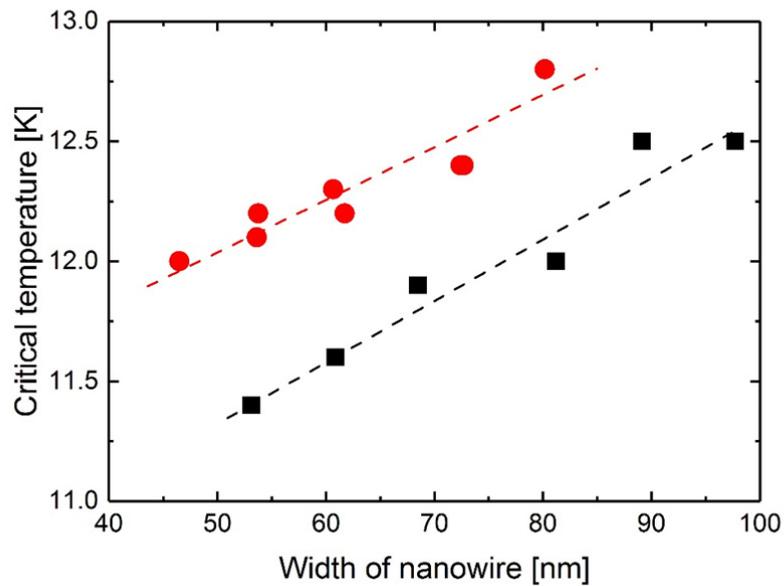

FIG. 2 Dependence of the critical temperature on the width for nanowires made by the positive-PMMA (black squares) and the negative-PMMA (red circles) lithography. The dashed lines are to guide the eyes.

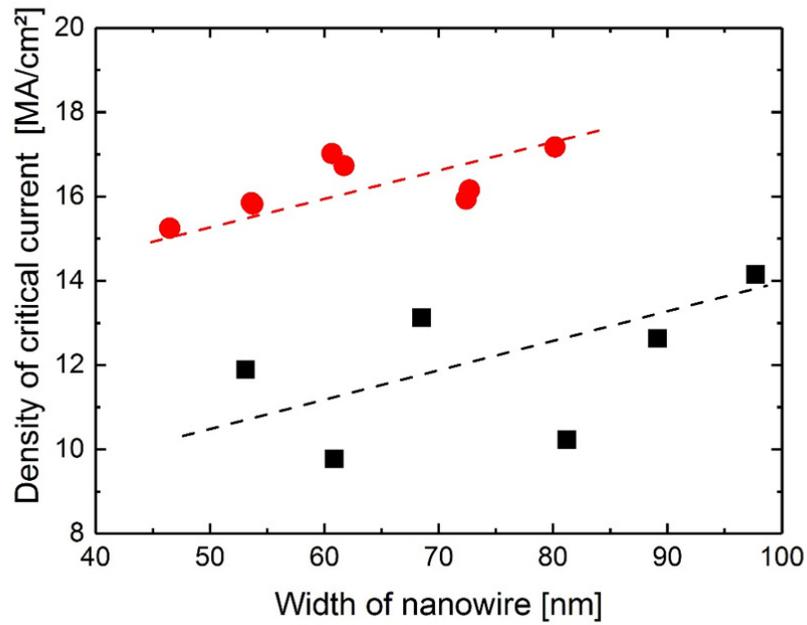

FIG. 3 Dependence of the critical current density at $T = 4.2$ K on the width for nanowires made by the positive-PMMA (black squares) and the negative-PMMA (red circles) lithography. The dashed lines are to guide the eyes.

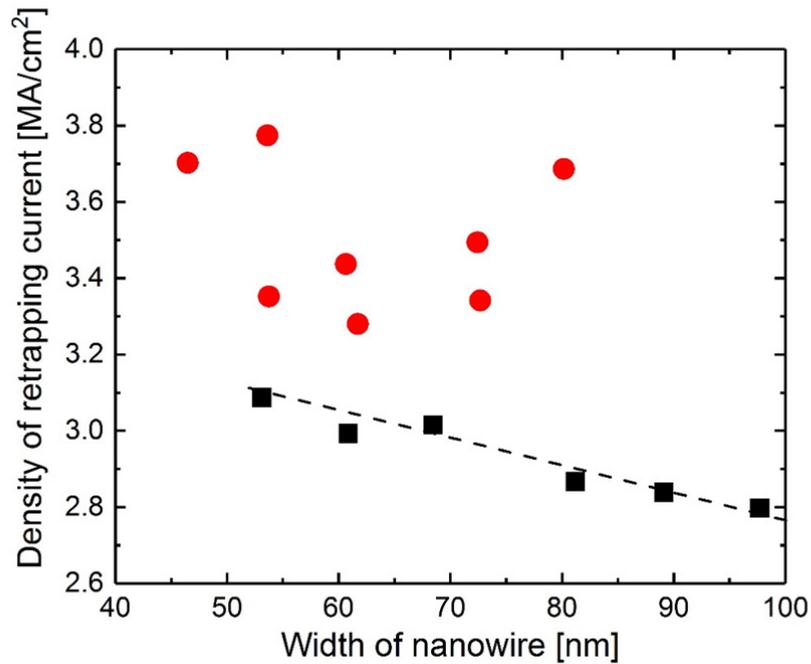

FIG. 4 Dependence of the retrapping current density at $T = 4.2$ K on the width for nanowires made by the positive-PMMA (black squares) and the negative-PMMA (red circles) lithography. The dashed line is to guide the eyes.

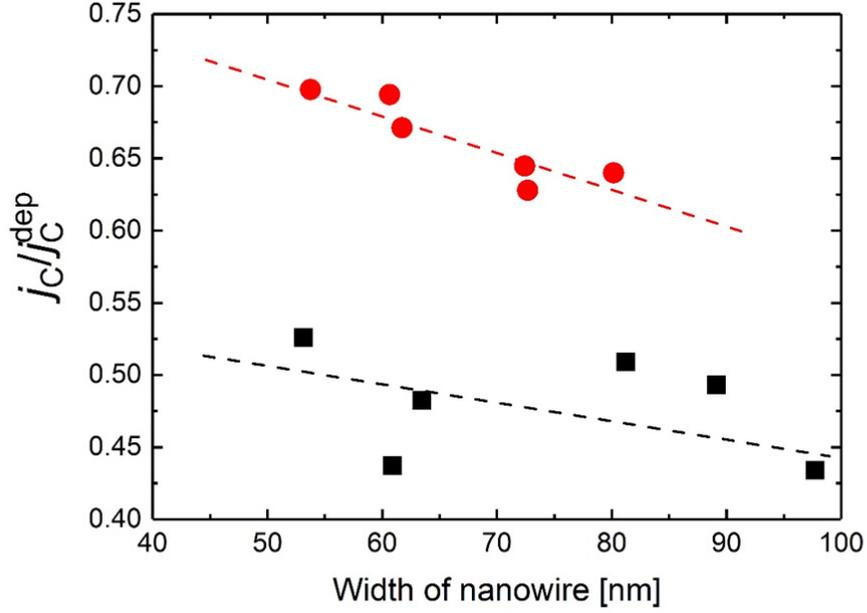

FIG. 5 Dependence of the ratio of the measured critical current density $j_C$ to the calculated (Eq. 3) density of the depairing current $j_C^{dep}$ (both at $T = 4.2$ K) on the width for nanowires made by the positive-PMMA (black squares) and the negative-PMMA (red circles) lithography. The dashed lines are to guide the eyes.

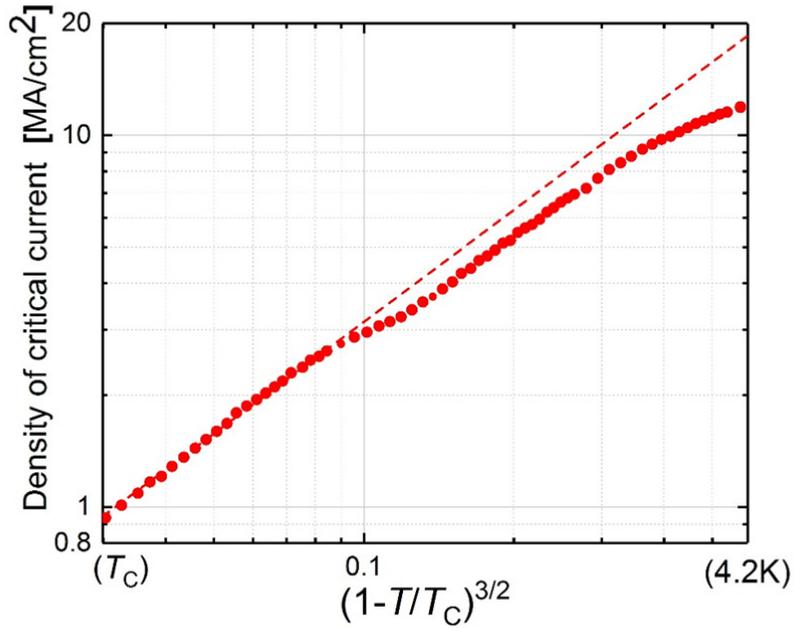

FIG. 6 Dependence of the density of the experimental critical current $j_C$ (red circles) on the reduced temperature $t = (1 - T/T_C)^{3/2}$ for the 80 nm-wide nanowire. The dashed line shows the best linear fit $j_C(t) = j_C^{extr}\, t$ of the experimental data near the transition temperature where $j_C^{extr}$ is the fit parameter.

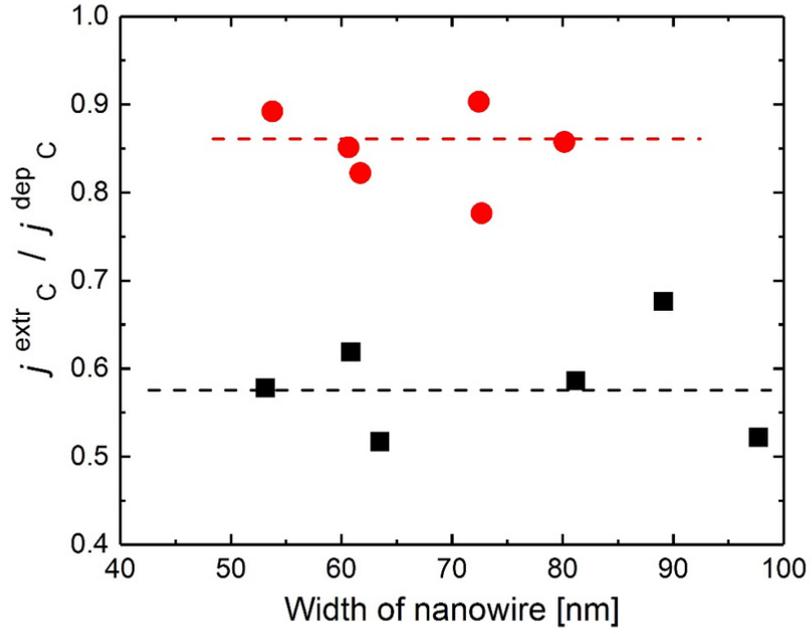

FIG. 7 Ratio of the extracted current density $j_C^{extr}$ (slope of the dashed line in Fig. 6) to the density of the depairing critical current $j_C^{dep}$ at $T = 0$ (Eq. 3) as function of nanowire width for nanowires made by the positive-PMMA (black squares) and the negative-PMMA (red circles) lithography. The dashed lines are to guide the eyes.